\documentclass[12pt]{article} 
\usepackage[T1]{fontenc} 
\usepackage[centertags]{amsmath} 
\usepackage{amsfonts} 
\usepackage{amssymb} 
\usepackage{amsthm} 
\usepackage{newlfont} 
\usepackage{epsfig} 
\usepackage{amscd}

\newcommand{\bd}{{\boldsymbol d}}

\newcommand{\be}{{\boldsymbol e}}

\newcommand{\bA}{{\boldsymbol A}}

\newcommand{\PP}{{\mathbb P}}

\newcommand{\ZZ}{{\mathbb Z}} 
\newcommand{\KK}{{\mathbb K}} 
\newcommand{\FF}{{\mathbb F}} 
\newcommand{\LL}{{\mathbb L}}

\newcommand{\calC}{{\mathcal C}} 

\newtheorem{Th}{Theorem}
\newtheorem{Cor}[Th]{Corollary} 
\newtheorem{Lem}[Th]{Lemma} 
\newtheorem{Prop}[Th]{Proposition} 
\theoremstyle{definition} 
\newtheorem{Def}{Definition}
\theoremstyle{remark} 
\newtheorem*{Rem}{Remark}
\begin{document} 
\begin{center} 
{\Large The discrete KP and KdV equations over finite fields} 
  
\bigskip 
 
{\large Mariusz Bia{\l}ecki$^{\dag\ddag\natural}$ and Adam Doliwa$^{*\natural}$}

\bigskip   
 
$^\dag${\it Instytut Fizyki Teoretycznej, Uniwersytet w Bia{\l}ymstoku 
 
ul. Lipowa 41, 15-424 Bia{\l}ystok, Poland}  
 
\bigskip 
 
$^\ddag${\it Instytut Geofizyki PAN 
 
ul. Ksi\c{e}cia Janusza 64, 01-452 Warszawa, Poland} 
 
\bigskip 
 
$^*${\it Wydzia{\l} Matematyki i Informatyki, Uniwersytet 
Warmi\'nsko--Mazurski 
 
ul. \.Zo{\l}nierska 14A, 10-561 Olsztyn, Poland} 

\bigskip 
 
$^\natural${\it Instytut Fizyki Teoretycznej, Uniwersytet 
Warszawski
 
ul. Ho\.{z}a 69, 00-681, Poland}

\end{center}

\begin{abstract} 
We propose the algebro-geometric method of construction of solutions 
of the discrete KP equation over a finite field. We also perform the 
corresponding reduction to the finite field version of discrete KdV 
equation. We write down formulas which allow to construct multisoliton
solutions of the equations starting from vacuum wave functions on 
arbitrary non-singular curve. 
 
\end{abstract} 
 
\section{Introduction}  
 
The goal of this paper is to present a general algebro-geometric
method of 
construction of solutions to the cellular automaton associated with the 
discrete Kadomtsev--Petviashvilii (KP) equation~\cite{Hirota}
\begin{multline*} 
\tau (n_1+1,n_2,n_3)\;\tau (n_1,n_2+1,n_3+1) -  
 \tau (n_1,n_2+1,n_3) \; \tau (n_1+1,n_2,n_3+1) + \\ 
+ \tau (n_1,n_2,n_3+1)\;\tau (n_1+1,n_2+1,n_3) =0,
\end{multline*} 
and with its reduction to the discrete Korteweg--de~Vries (KdV)
equation~\cite{HirotadKdV}
\begin{multline*} 
\tau (n_1+1,n_3)\;\tau (n_1,n_3) -  
 \tau (n_1,n_3-1) \; \tau (n_1+1,n_3+1) + \\ 
+ \tau (n_1,n_3+1)\;\tau (n_1+1,n_3-1) =0. 
\end{multline*} 
It turns out that main algebro-geometric ideas of construction of solutions
of the equations can be transferred from the level of Riemann surfaces to
the level of algebraic curves over finite fields.
Our motivation to extend validity of the discrete equations to finite field
domain was presented in the recent paper \cite{ffHirotaDBK},
where we also send for relevant literature. 
 
The layout of the paper is as follows. In Section~\ref{sec:alg-dKP} we 
present the general algebro-geometric scheme for construction of  
solutions of the discrete KP equation. In Section~\ref{sec:alg-dKdV} we 
describe the algebro-geometric reduction scheme from KP to
the discrete KdV equation.   
Section~\ref{sec:vacuum+N-soliton} is devoted to construction of 
multisoliton solutions (on nontrivial background)
starting from the vacuum 
wave functions on algebraic curves over finite fields. As a simple
application of the general method we construct multisolitonic solutions
starting from the algebraic curve of genus zero (the projective line).

\section{Solutions of the discrete KP equation from algebraic curves 
over finite fields} 
\label{sec:alg-dKP} 
This Section is motivated by algebro-geometric (over the complex field) 
approach to the discrete KP (or Hirota) equation (see for
example~\cite{Krichever-acnde} and \cite{KWZ}) and by \cite{ffHirotaDBK},
where an equivalent version of the discrete KP equation (the discrete
analogue of the Toda field system) was studied in detail in the context of
finite field valued solutions.

Consider an algebraic projective curve $\calC$,  
absolutely irreducible, nonsingular, of genus $g$, defined over the finite  
field $\KK=\FF_q$ with $q$ elements, where $q$ is a power of a 
prime integer $p$ (see, for example \cite{Stichtenoth}). 
By $\calC_{\KK}$ we denote 
the set of $\KK$-rational points of the curve. 
By $\overline{\KK}$ denote the algebraic closure of  
$\KK$, i.e., $\overline{\KK} = \bigcup_{\ell=1}^\infty \FF_{q^\ell}$, and by 
$\calC_{\overline{\KK}}$ denote the corresponding (infinite) set of 
$\overline{\KK}$-rational points of the curve.  
The action of the Galois group $G(\overline{\KK}/\KK)$ (of automorphisms of  
$\overline{\KK}$ which are identity on $\KK$, see \cite{Lang-alg}) extends 
naturally to action on $\calC_{\overline{\KK}}$. 
 
Let us choose:\\ 
1. three points $a_i\in\calC_\KK$, \  $i=1,2,3$, \\ 
2. $N$ points $c_\alpha\in\calC_{\overline{\KK}}$, $\alpha=1,\dots,N$, 
which satisfy the following $\KK$-rationality condition 
\[ 
\forall \sigma\in  G(\overline{\KK}/\KK), \quad  
\sigma(c_\alpha) = c_{\alpha^\prime}, 
\] 
3. $N$ pairs of points $d_\beta, e_\beta\in\calC_{\overline{\KK}}$,  
$\beta=1,\dots,N$, which satisfy the following $\KK$-rationality condition  
\begin{equation*} \label{eq:K-rat-cond} 
\forall \sigma\in  G(\overline{\KK}/\KK): \quad 
\sigma(\{d_\beta,e_\beta\})=\{d_{\beta^\prime},e_{\beta^\prime}\}, 
\end{equation*} 
4. $g$ points $f_\gamma\in\calC_{\overline{\KK}}$, $\gamma=1,\dots,g$, 
which satisfy the following $\KK$-rationality condition 
\[ 
\forall \sigma\in  G(\overline{\KK}/\KK), \quad  
\sigma(f_\gamma) = f_{\gamma^\prime}, 
\] 
5. the normalization point $a_0\in\calC_\KK$. 

As a rule we consider here only the generic case and assume that all the 
points used 
in the construction are generic and distinct. In particular, 
the divisor $D=\sum_{\gamma=1}^g f_\gamma$ is non-special. 
We remark that it is enough the check the $\KK$-rationality conditions  
in any extension field $\LL\supset\KK$   
of rationality of all the points used in the construction. 
 
\begin{Def} 
Fix $\KK$-rational local parameters $t_i$ at $a_i$, $i=0,1,2,3$. 
For any integers $n_1,n_2,n_3\in \ZZ$ define the wave function  
$\psi(n_1,n_2,n_3)$ as a 
rational function with the following properties\\ 
1. it has pole of the order at most $n_1+n_2+n_3$ at $a_0$,\\  
2. the first nontrivial coefficient of its expansion in $t_0$ at $a_0$ is 
normalized to one, \\  
3. it has zeros of order at least $n_i$ at $a_i$ for $i=1,2,3$, \\  
4. it has at most simple poles at points $c_\alpha$, $\alpha=1,\dots,N$, \\ 
5. it has at most simple poles at points $f_\gamma$, $\gamma=1,\dots,g$,\\ 
6. it satisfies $N$ constraints 
\begin{equation*} \label{eq:constraints} 
\psi(n_1,n_2,n_3)(d_\beta)=\psi(n_1,n_2,n_3)(e_\beta), \quad 
\beta=1,\dots,N. 
\end{equation*} 
\end{Def} 
The function $\psi(n_1,n_2,n_3)$ is  $\KK$-rational, which follows  
from $\KK$-rationality conditions of sets of points in their definition. 
As usual, zero (pole) of a negative order means pole (zero) of the 
corresponding positive order. Correspondingly one should exchange the 
expressions "at most" and "at least" in front of the orders of poles and 
zeros. 
By the standard application of the Riemann--Roch theorem (and the genericity
assumption) we conclude that  
the wave function $\psi(n_1,n_2,n_3)$ exists and is unique. 
 
In the generic case, which we assume in the sequel, when the order of  
the pole of $\psi$ at $a_0$ is 
$(n_1+n_2+n_3)$ denote by $\zeta^{(0)}_k(n_1,n_2,n_3)$  
and $\zeta^{(i)}_k(n_1,n_2,n_3)$, $i=1,2,3$, 
$\KK$-rational
coefficients of expansion of $\psi$ at $a_0$ and at $a_i$, respectively, 
i.e.,
\[ \psi = \frac{1}{t_0^{(n_1+n_2+n_3)}} 
\left( 1 + \sum_{k=1}^{\infty} \zeta^{(0)}_k  
t_0^k \right), \qquad 
\psi = t_i^{n_i} \sum_{k=0}^{\infty} \zeta^{(i)}_k t_i^k .  
\] 
Denote by $T_i$ the operator of translation 
in the variable $n_i$, $i=1,2,3$, 
for example $T_1 \psi(n_1,n_2,n_3) = \psi(n_1+1,n_2,n_3) $. 
Uniqueness of the wave function implies the following statement. 
\begin{Prop} \label{prop:equations-psi} 
The function $\psi$  
satisfies equations 
\begin{equation} 
T_i \psi - T_j \psi +  
\frac{T_j\zeta^{(i)}_ 0}{\zeta^{(i)}_0} \psi = 0 , \quad i\ne j. 
\label{eq:psi}  
\end{equation} 
\end{Prop} 
Notice that equation \eqref{eq:psi} gives
\begin{equation} \label{eq:zeta-zeta}
\frac{T_j\zeta^{(i)}_ 0}{\zeta^{(i)}_0} = -  
\frac{T_i\zeta^{(j)}_ 0}{\zeta^{(j)}_0} , \quad i\ne j.  
\end{equation}
Define $ \rho_i= (-1)^{\sum_{j<i} n_j} \zeta^{(i)}_0$, then equation
\eqref{eq:zeta-zeta} implies existence of a $\KK$-valued potential 
(the $\tau$-function) defined (up to a multiplicative
constant) by formulas 
\begin{equation} \label{eq:tau-def} 
 \frac{T_i\tau}{\tau} = \rho_i,\quad  i=1,2,3.  
\end{equation} 
Finally, equations \eqref{eq:psi} give rise to condition 
\begin{equation*}  \label{eq:KP-rho}
\frac{T_2\rho_1}{\rho_1} - \frac{T_3\rho_1}{\rho_1} + 
\frac{T_3\rho_2}{\rho_2} = 0, 
\end{equation*}  
which written in terms of the $\tau$-function gives the discrete KP
equation~\cite{Hirota}
\begin{equation}  \label{eq:tauKP} 
(T_1\tau) \;(T_2T_3\tau) - (T_2\tau) \;(T_3T_1 \tau) + 
(T_3\tau) \;(T_1T_2\tau) = 0. 
\end{equation} 
\begin{Cor} \label{cor:symmetry-tau}
Notice that multiplication of $\rho_i$, $i=1,2,3$, by a function of the
single argument $n_i$, do not affects nor existence of the $\tau$ function
nor equation \eqref{eq:tauKP} satisfied by the function. 
\end{Cor}

\section{Reduction to the discrete KdV equation}
\label{sec:alg-dKdV}
The discrete KdV equation \cite{HirotadKdV,KWZ}
\begin{equation}  \label{eq:tauKdV} 
(T_1\tau) \;\tau - (T_3^{-1}\tau) \;(T_3T_1 \tau) + 
(T_3\tau) \;(T_1T_3^{-1}\tau) = 0, 
\end{equation} 
is obtained from the discrete KP equation by imposing
constraint
\begin{equation} \label{eq:KPtoKdVconstr}
T_2T_3 \tau = \gamma \tau,
\end{equation}
where $\gamma$ is a non-zero constant.
Transition of the reduction \eqref{eq:KPtoKdVconstr} to the level of the 
wave function $\psi$ will give the algebro-geometric procedure of 
construction of solutions of the discrete KdV equation. 
\begin{Lem} \label{lem:KP-to-KdV}
Assume that on the algebraic curve $\calC$ there exists a meromorphic
function $h$ with the following properties\\
1. it has two simple zeroes at points $a_2$ and $a_3$ and no
other zeroes,\\ 
2. it has double pole at $a_0$,\\
3. it satisfies $N$ constraints $h(d_\beta)=h(e_\beta)$,
$\beta=1,\dots,N$,\\
4. the first nontrivial coefficient of 
its expansion in the parameter $t_0$ at $a_0$ is normalized to one.\\
Then the wave function $\psi$ satisfies the following condition
\begin{equation} \label{eq:KPtoKdV-constr-psi}
T_2T_3\psi = h \psi.
\end{equation}
\end{Lem}
\begin{Rem}
Existence of such a function $h$ implies that the algebraic curve $\calC$ is
hyperelliptic.
\end{Rem}
\begin{Prop} \label{prop:KP-to-KdV}
Let $h$ be the function as in Lemma~\ref{prop:KP-to-KdV}. Assume
additionally that
\begin{equation} \label{eq:assumpt-ha1}
h(a_1)=1.
\end{equation} 
Denote by $\delta_2$ and $\delta_3$ the first 
coefficients of local
expansion of $h$ in parameters $t_2$ and $t_3$ at $a_2$ and $a_3$, 
correspondingly
\begin{equation*}
h=t_2(\delta_2 + \dots), \qquad    h=t_3(\delta_3 + \dots).
\end{equation*}
Then the function
\begin{equation} \label{eq:tau-tau-tylda}
\tilde\tau= \tau \, \delta_2^{-n_2(n_2-1)/2}(-\delta_3)^{-n_3(n_3-1)/2}
\end{equation}
satisfies the discrete KdV equation \eqref{eq:tauKdV}.
\end{Prop}
\begin{proof}
Expanding equation \eqref{eq:KPtoKdV-constr-psi} at $a_1$ and using of the
additional assumption \eqref{eq:assumpt-ha1} we obtain that
\begin{equation*} \label{eq:KPtoKdV-constr-rho-1}
T_2T_3\rho_1= \rho_1.
\end{equation*}
Expansions of equation \eqref{eq:KPtoKdV-constr-psi} at $a_2$ and $a_3$ 
give
\begin{equation*}
T_2T_3 \rho_2 = \delta_2\rho_2, \qquad  
T_2T_3 \rho_3 = -\delta_3\rho_3.
\end{equation*}
Therefore the functions 
\begin{equation*} \label{eq:rho-rho-tylda}
\tilde\rho_1=\rho_1, \qquad \tilde\rho_2=\delta_2^{-n_2}\rho_2, \qquad
\tilde\rho_3=(-\delta_3)^{-n_3}\rho_3,
\end{equation*} 
satisfy condition
\begin{equation} \label{eq:KPtoKdV-constr-rho-i}
T_2T_3\tilde\rho_i= \tilde\rho_i , \qquad  i=1,2,3.
\end{equation}
By Corollary~\ref{cor:symmetry-tau} the functions $\tilde\rho_i$, 
$i=1,2,3$, define new potential $\tilde\tau$, connected
with $\tau$ by \eqref{eq:tau-tau-tylda},
which satisfies the discrete KP equation \eqref{eq:tauKP}.
Moreover, conditions \eqref{eq:KPtoKdV-constr-rho-i} imply
that $\tilde\tau$ is subjected to constraint \eqref{eq:KPtoKdVconstr}.
\end{proof}
\begin{Rem}
Notice that in the above procedure one obtains a family, labelled by the
parameter $n_2$, of solutions of the discrete KdV equation.
\end{Rem}

\section{Construction of solutions of the discrete KP and KdV equations
using vacuum functions} 
\label{sec:vacuum+N-soliton} 
In this Section we write down results which allow to construct the
$N$-soliton $\tau$-function
starting from vacuum ($N=0$) wave functions
on algebraic curve over a finite field. The methods to obtain 
these results are the same as in the corresponding section 
of~\cite{ffHirotaDBK}.

In the case $N=0$ let us add superscript $0$ to all functions defined  
above. Define auxiliary vacuum wave functions $\phi^0_{\alpha}$, 
$\alpha=1,\dots,N$, as follows. 
\begin{Def} 
Fix local parameters $t_\alpha$ at $c_\alpha$, $\alpha=1,\dots,N$. 
For any $\alpha$ define the function $\phi^0_{\alpha}$ by the following  
set of conditions:\\ 
1. it has pole of the order at most $n_1+n_2+n_3-1$ at $a_0$,  \\ 
2. it has zeros of order at least $n_i$ at $a_i$, for $i=1,2,3$, \\  
3. it has at most simple pole at the point $c_\alpha$,\\ 
4. the first nontrivial  
coefficient of its expansion in $t_\alpha$ at $c_\alpha$ is 
normalized to one, \\ 
5. it has at most simple poles at points $f_\gamma$, $\gamma=1,\dots,g$. 
\end{Def} 
Using the Riemann-Roch theorem it can be shown that the function 
$\phi^0_{\alpha}$ exists and is unique. 
\begin{Prop} 
Denote by $\psi^0(\bd,\be)$, the column with $N$ entries  
of the form 
\[ 
\left[\psi^0(\bd,\be)\right]_\beta = 
\psi^0(d_\beta)-\psi^0(e_\beta), \quad \beta=1,\dots,N,  
\] 
denote by $\phi^0_{\bA}$  
the row with $N$ entries  
\[ \left[ \phi^0_{\bA} \right]_\alpha = \phi^0_{\alpha}, \quad 
\alpha=1,\dots,N,  
\] 
and denote by $\phi^0_{\bA,m}(\bd,\be)$ the 
$N\times N$ matrix whose element in row $\beta$ and  
column $\alpha$ is 
\[ 
\left[\phi^0_{\bA}(\bd,\be)\right]_{\alpha \beta}= 
\phi^0_{\alpha}(d_\beta)-\phi^0_{\alpha}(e_\beta) , \quad 
\alpha,\beta=1,\dots,N. 
\] 
Then the wave function $\psi$ of the discrete KP equation reads 
\begin{equation*} \label{eq:psi-i-N-soliton} 
\psi = \psi^0 - \phi^0_{\bA}  
[\phi^0_{\bA}(\bd,\be)]^{-1} 
\psi^0(\bd,\be). 
\end{equation*}      
\end{Prop} 
In the generic case denote by $H^0_{0,\alpha}$ the first nontrivial  
coefficient of  
expansion of the function $\phi^0_{\alpha}$ at $a_0$ in the  
uniformization parameter $t_0$, 
\[ \phi^0_{\alpha} = \frac{1}{t_0^{(n_1+n_2+n_3-1)}}\left( H^0_{0,\alpha} 
 + \dots 
\right). 
\]  
\begin{Cor} 
The corresponding expressions for  $\rho_{i}$ read 
 \begin{equation*} \label{eq:rho-N-soliton} 
\rho_{i}  =  \rho_{i}^0\left( 1 + (T_iH^0_{0,\bA})  
[\phi^0_{\bA}(\bd,\be)]^{-1} \psi^0(\bd,\be) \right), \qquad i=1,2,3,
\end{equation*} 
 where $H^0_{0,\bA}$ is 
the row with $N$ entries $H^0_{0,\alpha}$. 
\end{Cor} 
\begin{Prop} \label{prop:tau-N-soliton} 
The $\tau$-function can be constructed by the 
following formula 
\begin{equation} \label{eq:tau-N-soliton} 
\tau = \tau^0 \det \phi^0_{\bA}(\bd,\be). 
\end{equation}      
\end{Prop} 
\begin{Cor} 
Starting with $\KK$-valued function $\tau^0$ and the local 
parameters $t_\alpha$ at $c_\alpha$ chosen in a consistent way with the 
action of the Galois group $G(\overline{\KK}/\KK)$  
on $\calC_{\overline{\KK}}$ we obtain $\KK$-valued function $\tau$.  
\end{Cor} 
\begin{Cor} \label{cor:tau-KP-KdV-from-vac}
Notice that equation \eqref{eq:tau-tau-tylda} implies that the same formula
\eqref{eq:tau-N-soliton} holds also for the $\tilde\tau$-function in the
reduction from the discrete KP equation to the discrete KdV equation. 
\end{Cor}

We present here explicit formulas for the vacuum functions in the simplest 
case $g=0$. In constructing the vacuum functions 
we will use the standard parameter $t$ on the  
projective line $\PP(\KK)$ and we put $a_0=\infty$. 
Explicit form of the vacuum wave function reads 
\begin{eqnarray} 
\psi^0 &= & 
{(t-a_1)^{n_1}} 
{(t-a_2)^{n_2}} {(t-a_3)^{n_3}} \nonumber 
\end{eqnarray}     
which gives formulas for the functions $\rho_1$, $\rho_2$ and $\rho_3$ 
\begin{eqnarray} 
 \rho_1^0 &= & {(a_1-a_2)^{n_2}}{(a_1-a_3)^{n_3}}, \nonumber \\ 
 \rho_2^0 &=& {(-1)^{n_1}} {(a_2-a_1)^{n_1}}{(a_2-a_3)^{n_3}} , 
 \nonumber \\ 
 \rho_3^0 &=& {(-1)^{n_1+n_2}} {(a_3-a_1)^{n_1}}{(a_3-a_2)^{n_2}} . 
 \nonumber 
\end{eqnarray}     
Explicit form of the vacuum $\tau$-function reads 
\[ 
\tau^0 = {(a_1-a_2)^{n_1 n_2}} {(a_1-a_3)^{n_1 n_3}} {(a_2-a_3)^{n_2 n_3}}.    
\] 
The auxiliary vacuum wave functions $\phi^0_{\alpha}$, $\alpha=1,\dots,N$ 
have the form 
\begin{equation} \label{eq:phi-KP-vac} 
\phi_{\alpha}^0= \frac{1}{t-c_\alpha} \cdot 
\frac{(t-a_1)^{n_1} (t-a_2)^{n_2} (t-a_3)^{n_3}} 
{(c_\alpha-a_1)^{n_1} (c_\alpha-a_2)^{n_2}(c_\alpha-a_3)^{n_3}}. 
\end{equation} 

In the case of the reduction from KP to KdV the function $h$ reads
\begin{equation*}
h(t) = (t-a_2)(t-a_3),
\end{equation*}
while the points $a_i$, $i=1,2,3$, are subjected to the condition
\begin{equation*} \label{eq:constr-h-vac}
h(a_1)=(a_1-a_2)(a_1-a_3)=1.
\end{equation*}
Then, in notation of Section~\ref{sec:alg-dKdV},
\begin{equation*}
\delta_2 = a_2-a_3 = -\delta_3,
\end{equation*}
and, according to Proposition~\ref{prop:KP-to-KdV}, the vacuum solution of
the discrete KdV equation reads
\begin{equation} \label{eq:phi-KdV-vac}
\tilde\tau^0=(a_1-a_3)^{n_1(n_3-n_2)}(a_2-a_3)^{n_2-(n_3-n_2)(n_3-n_2-1)/2}.
\end{equation}  
Finally, by Corollary \ref{cor:tau-KP-KdV-from-vac},
formulas \eqref{eq:tau-N-soliton}--\eqref{eq:phi-KdV-vac} allow
to find pure $N$-soliton solutions of the discrete KdV equation over finite
fields.

Let us present an example of finite field valued solution of the discrete
KdV equation in bilinear form. We take $\KK=\FF_5$, and all the points
used in the construction will be in $\FF_{5^2}$, which we consider as
extension of $\FF_5$ by the polynomial $w(x)=x^2 + x + 1$. 
The corresponding Galois group reads $G(\FF_{5^2}/\FF_5)=\{ id, \sigma\}$, 
where $\sigma$ is the Frobenius automorphism \cite{Lang-alg}. The
parameters of the solution are chosen as follows: 

$a_1=(00)$, $a_2=(02)$, $a_3=(03)$, 

$c_1=(10)$, $c_2=\sigma(c_1)=(44)$,

$d_1=(21)$, $e_1=\sigma(d_1)=(34)$, 

$d_2=(13)$, $e_2=\sigma(d_2)=(42)$.

\begin{figure}
\begin{center}
\leavevmode\epsfysize=6.2cm\epsffile{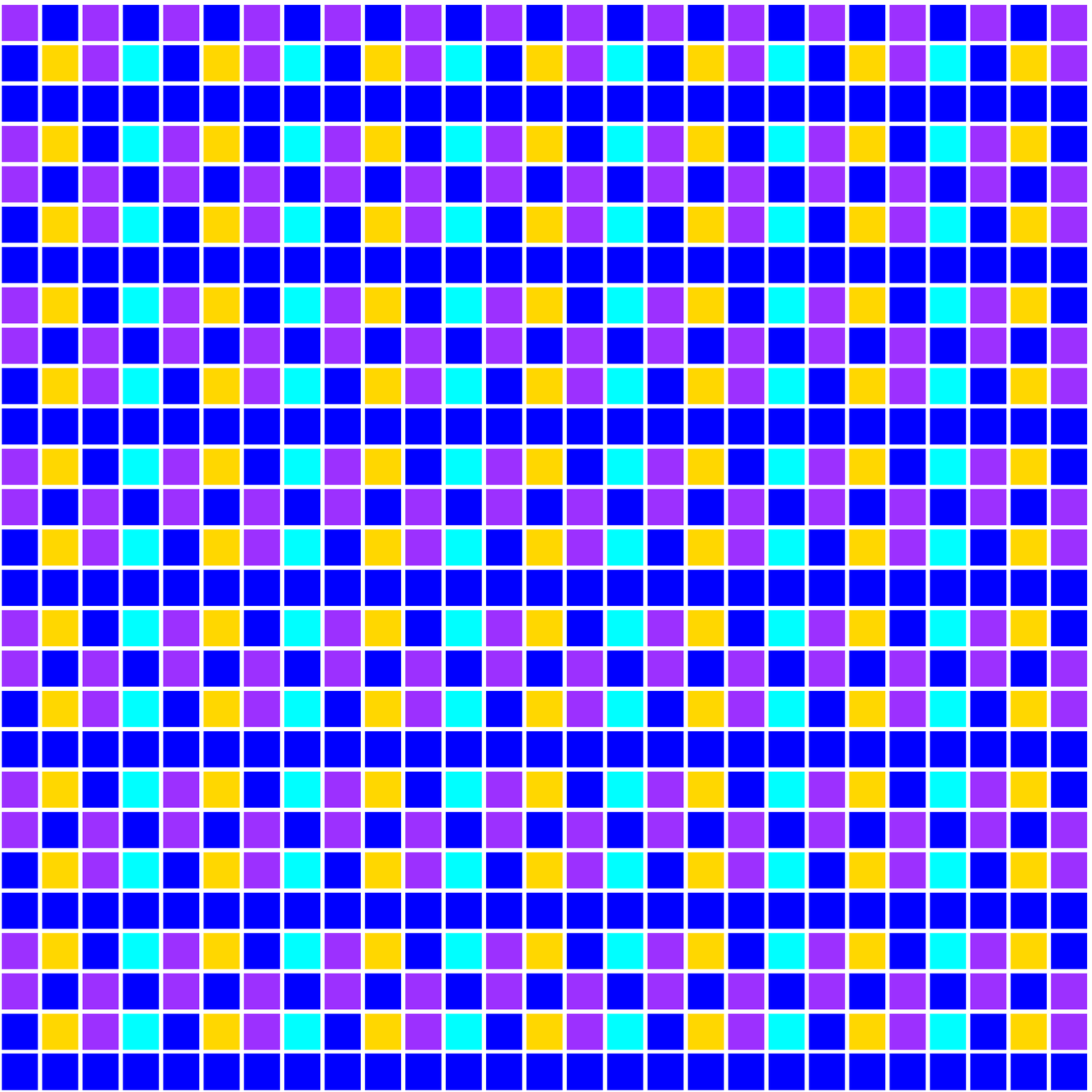} \hspace{0.5cm}
\leavevmode\epsfysize=6.2cm\epsffile{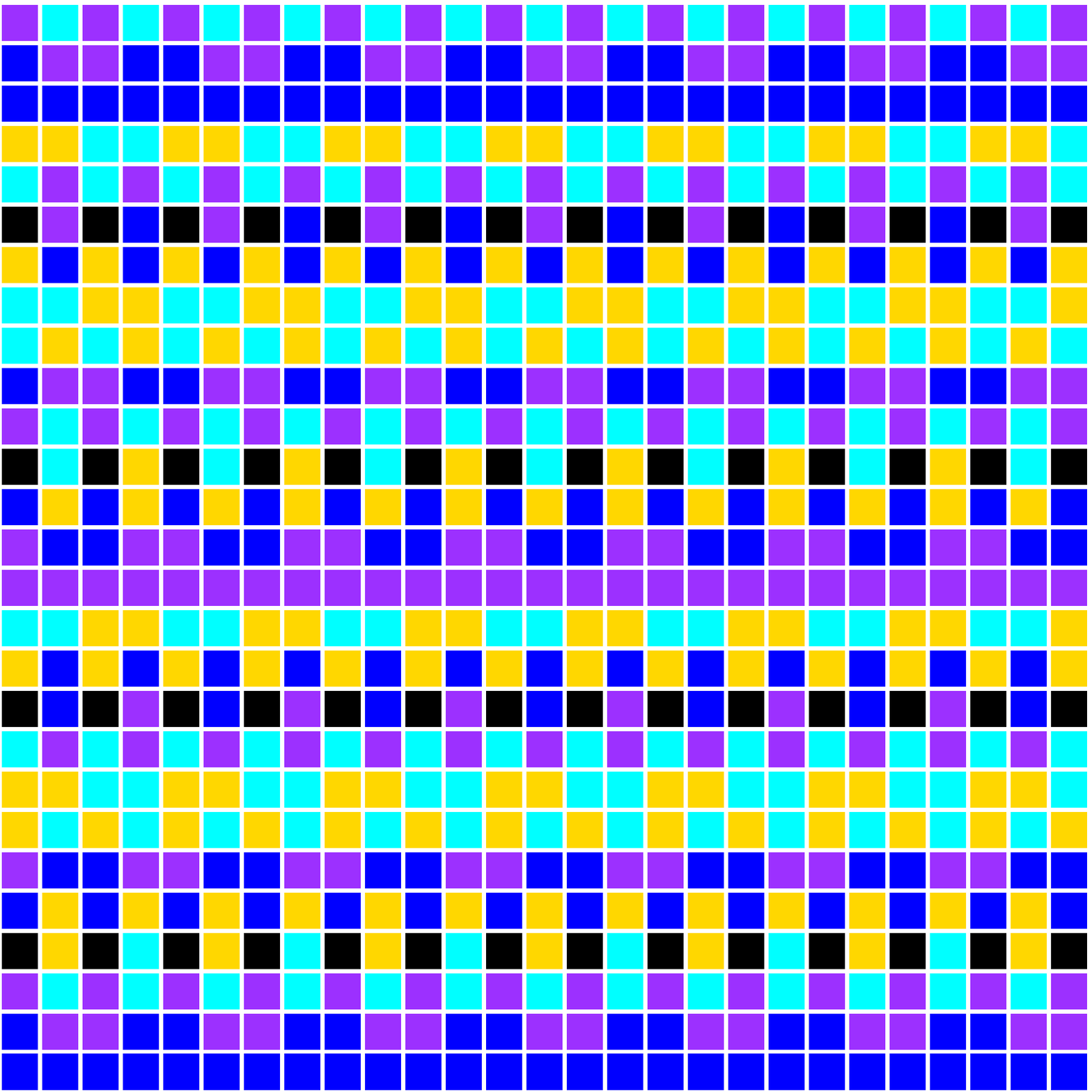}
\end{center}
\caption{
The vacuum and 2-soliton solutions of the discrete KdV equation in $\FF_5$; 
$n_1$ range from $0$ to $26$ (directed to the 
right), $n_3$ range from $0$ to $26$ (directed up).
}
\label{fig:2bKdV}
\end{figure}

The function $\tilde\tau$ is normalized to one for $n_1=n_2=n_3=0$. 
This solution of the discrete KdV equation, for $n_2=0$, is presented  
in comparison with the vacuum solution in Figure~\ref{fig:2bKdV}. 
The elements of $\FF_5$ are represented by:\\
\leavevmode\epsfysize=0.35cm\epsffile{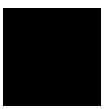} -- $(00)$, 
\leavevmode\epsfysize=0.35cm\epsffile{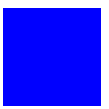} -- $(01)$, 
\leavevmode\epsfysize=0.35cm\epsffile{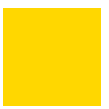} -- $(02)$,
\leavevmode\epsfysize=0.35cm\epsffile{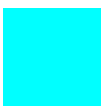} -- $(03)$, 
\leavevmode\epsfysize=0.35cm\epsffile{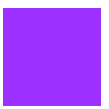} -- $(04)$.
The periods in variables $n_1$, $n_3$ are
$4$, $24$, correspondingly (both must be divisors of $24=|\FF^*_{5^2}|$, 
see~\cite{ffHirotaDBK}). 

Finally, we would like to remark that examples of multisoliton solutions on
curves of non-zero genus are more involved and their construction
needs some techniques on Jacobians of algebraic curves~\cite{BialDol-hyp}.

\section*{Acknowledgments} 
The authors wish to express their thanks to Pawe{\l} Klimczewski for
providing them with a computer program to visualize the solutions. 
M.~B. would like to thank the organizers of the NEEDS'02 meeting for 
support. The paper was partially supported by KBN grant no. 2P03B12622. 
 
\bibliographystyle{amsplain} 
\providecommand{\bysame}{\leavevmode\hbox to3em{\hrulefill}\thinspace}

\end{document}